\documentclass[conference]{IEEEtran}

\usepackage{siunitx}
\usepackage{graphicx}
\usepackage{booktabs}
\usepackage{tikz}
\usepackage{pgfplots}

\usepackage{hyperref}

\pgfplotsset{compat=1.18}

\graphicspath{fig}

\title{NeCTAr: A Heterogeneous RISC-V SoC for Language Model Inference in Intel 16}
\author{
\IEEEauthorblockN{
\parbox{0.99\textwidth}{
\centering
Viansa Schmulbach, 
Jason Kim, 
Ethan Gao,
Lucy Revina,
Nikhil Jha, 
Ethan Wu,  
Borivoje Nikoli\`c
}
}
\vspace{1ex}
\IEEEauthorblockA{University of California, Berkeley, CA, USA, bora@eecs.berkeley.edu}
}

\begin{document}

\maketitle

\begin{abstract}

This paper introduces NeCTAr (Near-Cache Transformer Accelerator), a 16nm heterogeneous multicore RISC-V SoC for sparse and dense machine learning kernels with both near-core and near-memory accelerators. A prototype chip runs at 400MHz at 0.85V and performs matrix-vector multiplications with 109 GOPs/W. The effectiveness of the design is demonstrated by running inference on a sparse language model, ReLU-Llama.

\end{abstract}

\section{Introduction}

Evolving machine learning workloads, such as those associated with large language model (LLM) inference, running on edge devices require heterogeneous specialized accelerators for efficient execution of both dense and sparse matrix kernels. Executing memory-bound dense matrix operations in near-memory accelerators avoids data movement and enables high-performance and low-energy inference on modern dense neural networks (DNNs). When implemented near the last-level cache, dense matrix engines can take advantage of high-bandwidth banked memories \cite{chen-cnc}.  For ease of programming and operating system support, the cache system must be coherent and backed by a virtual memory system. On the other hand, sparse kernels incur more control overhead in traversing the sparse data structures, so sparse accelerators should be implemented near the general-purpose processor cores.  By combining the near-memory accelerators with tightly coupled near-core sparse matrix engines, a broad range of modern ML workloads can efficiently be targeted.  

\begin{figure}
    \centering
    \includegraphics[width=0.9\columnwidth]{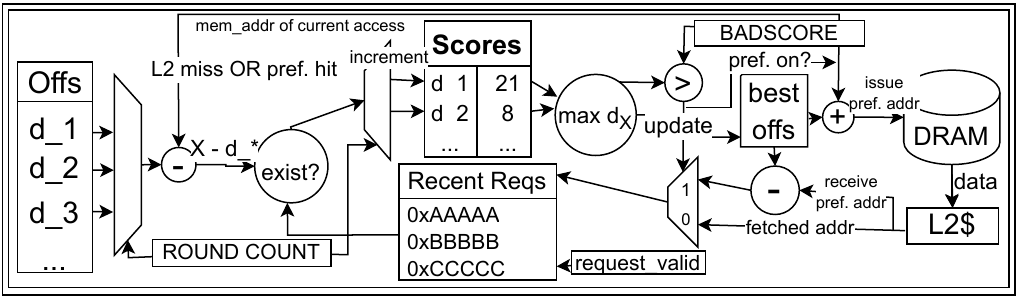}
    \caption{Best-offset prefetcher block diagram.}
    \label{fig:bop}
\end{figure}

This work presents a heterogeneous multicore SoC for ML inference that contains four in-order RISC-V cores, a 64b-wide TileLink NoC \cite{constellation}, an on-chip 16KB scratchpad, 256KB of L2 cache, and an on-chip 10MHz-2.2GHz PLL provided by Intel\@.  Onboard accelerators include four near-memory compute engines (\autoref{fig:nmce-arch}), four sparse matrix coprocessors (\autoref{fig:spaccel-arch}), and a best-offset prefetcher (\autoref{fig:bop}). NeCTAr, developed from concept to chip tapeout in less than fifteen weeks, serves as a demonstration of agile chip design for domain-specific applications. The design leverages the meta-programming and parameter systems made possible by open-source module generators and physical design flow tools within the Chipyard framework, integrating open-source libraries with commercial EDA tools. NeCTAr is capable of running a full 1.7M ReLU Llama model, achieving a rate of 1.28 infs/s when using the near-memory compute engines, limited by off-chip memory bandwidth.  When running a multiplication kernel, our near-memory compute engines have a speedup of 100x over software implementation on a multi-core processor. NeCTAr's design methodology and bring-up procedure serve as a platform for demonstrating the adaptation of SoC features to rapidly evolving machine learning requirements. 

\begin{figure}
    \centering
    \includegraphics[width=0.95\columnwidth]{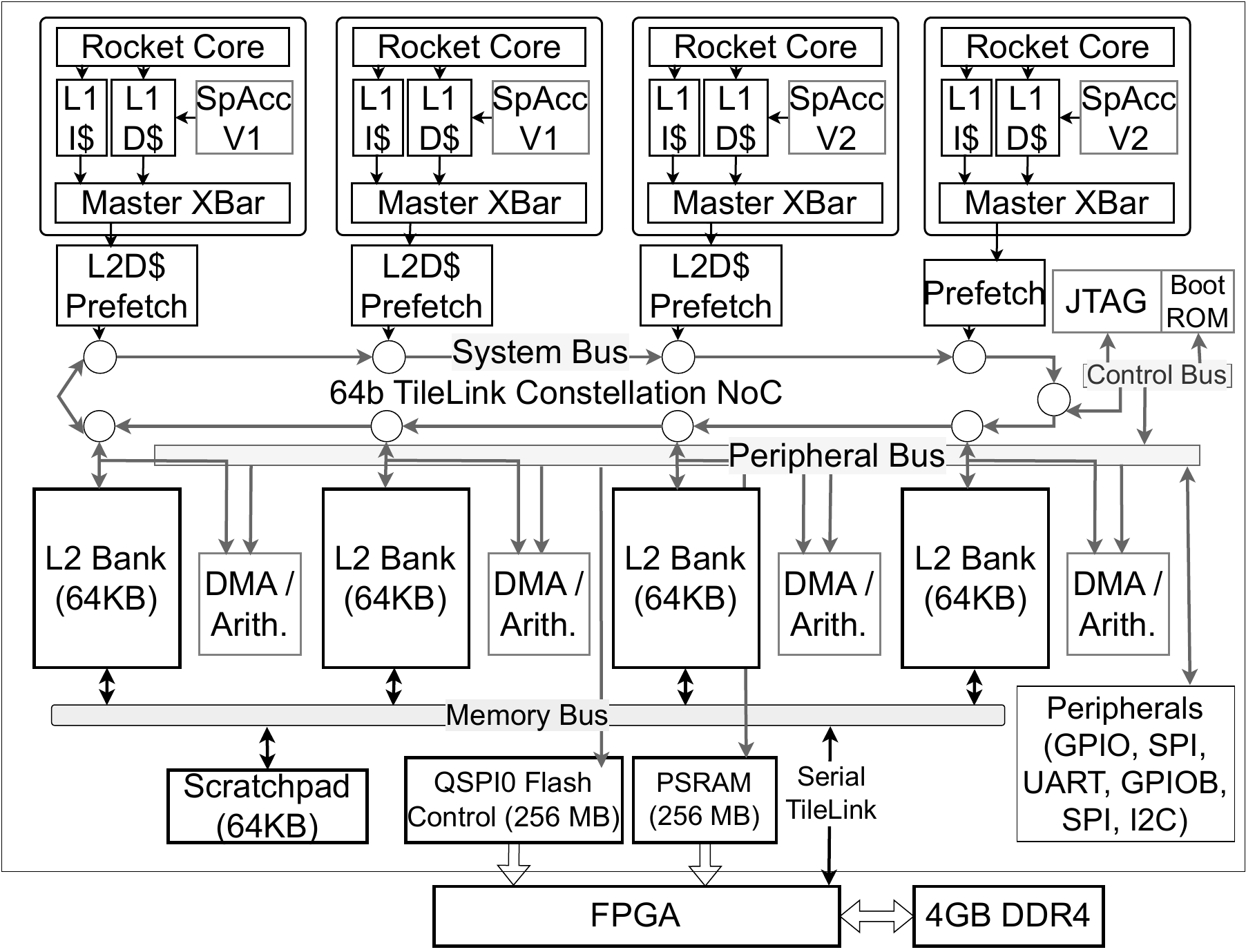}
    \caption{NeCTAr block diagram.}
    \label{fig:block-diagram}
\end{figure}

\begin{figure}
    \centering
    \parbox{0.55\columnwidth}{
        \footnotesize\centering
        \begin{tabular}{c|c}
            \multicolumn{2}{c}{Chip Specifications} \\ \midrule
            Technology & Intel 16 \\
            Area & \SI{4}{\square\milli\meter} \\
            SRAM & \SI{320}{\kilo\byte} \\
            Voltage & 0.55--1.15 \si{\volt} \\
            Freq & \SI{400}{\mega\hertz} \\
            Power & 7.3--171 \si{\milli\watt} \\
            Energy Eff. & \SI{108}{GOPs/\watt} \\
        \end{tabular}
    }
    \hfill
    \parbox{0.4\columnwidth}{
        \includegraphics[width=0.33\columnwidth]{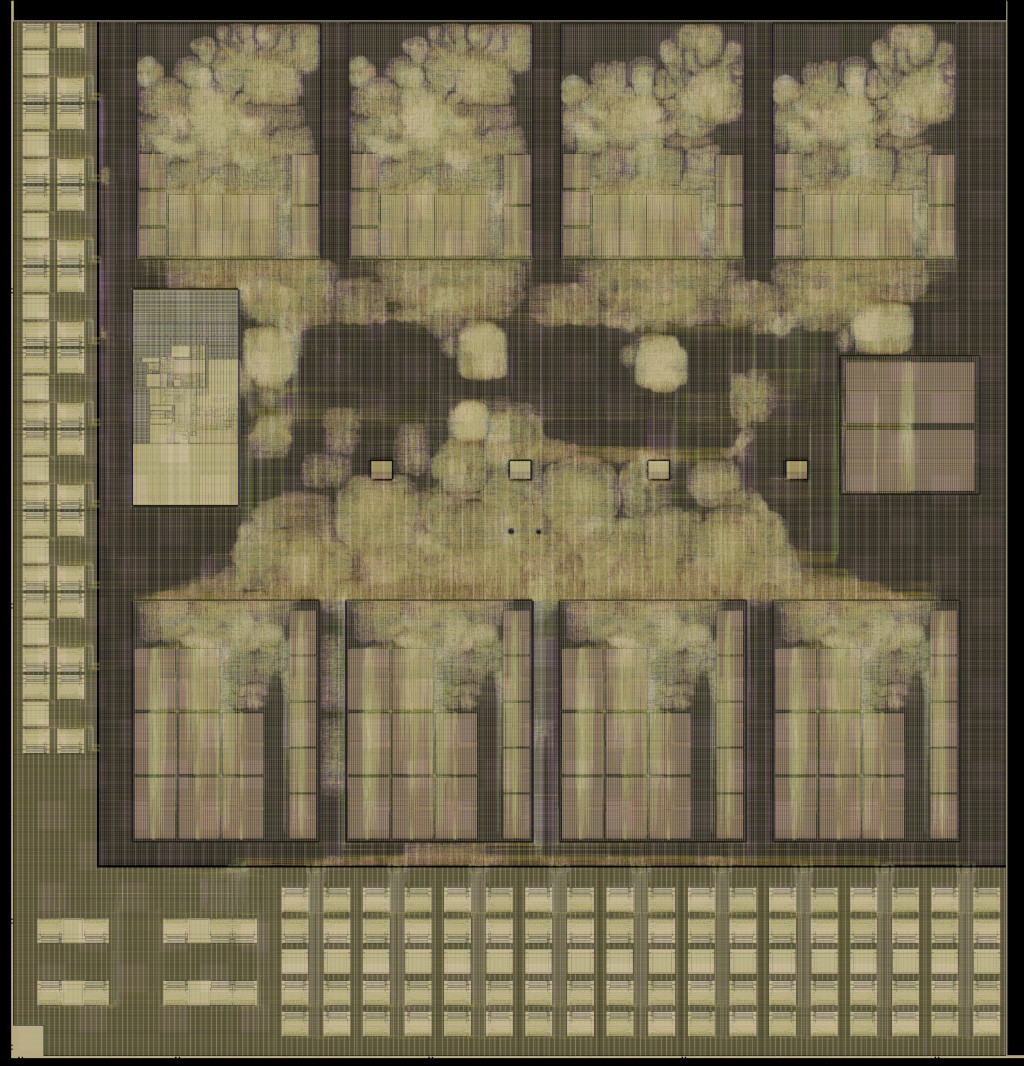}
    }
    \caption{Chip specifications and die.}
    \label{fig:chip-specifications}
\end{figure}

\section{Architecture and Implementation}

\subsection{Chip Architecture}

\begin{figure}
    \centering
    \hspace{-0.02\columnwidth}
    \includegraphics[width=1.05\columnwidth]{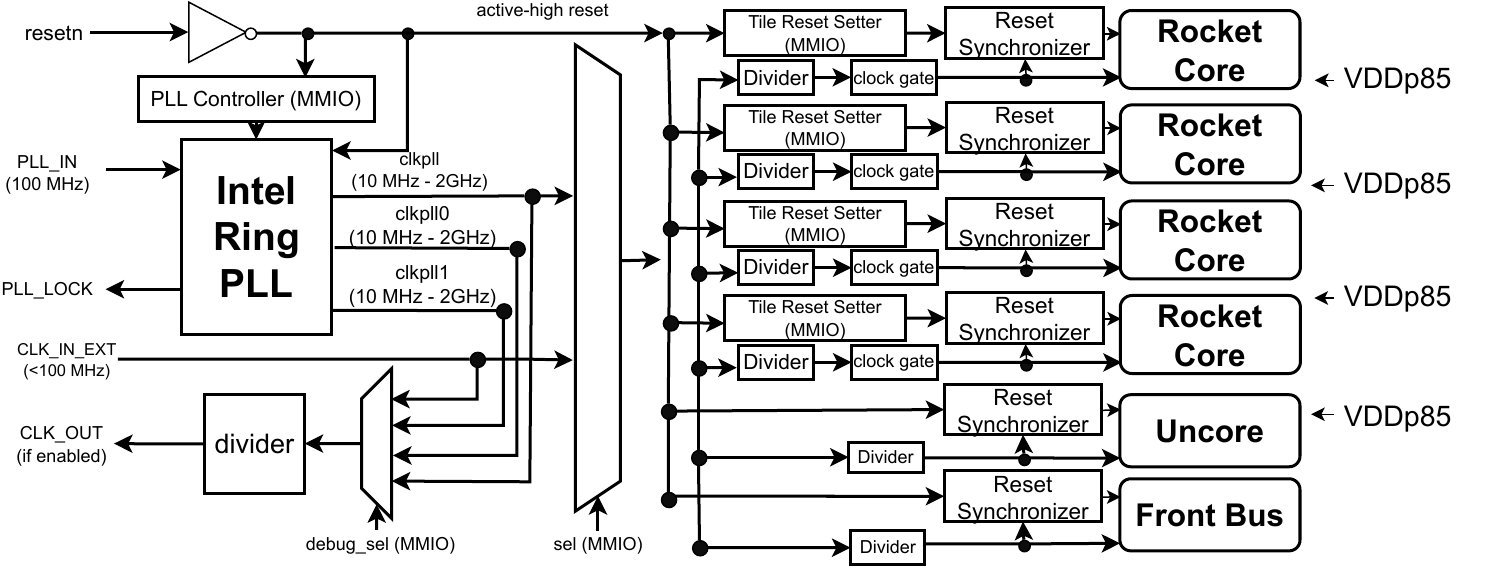}
    \caption{Clock tree diagram.}
    \label{fig:clocking}
\end{figure}

\autoref{fig:block-diagram} shows the chip’s architecture. Four 5-stage, in-order scalar RISC-V Rocket (RV64GC) cores \cite{chipyard} each have a tightly coupled sparse matrix accelerator and L2 prefetcher attached through a RoCC interface \cite{chipyard}. Chip interconnects are based on an updated Chipyard \cite{chipyard} architecture, with one primary ``system'' unidirectional torus NoC \cite{constellation}, a crossbar that attaches all peripherals, and a ``memory'' crossbar connecting different cached memory regions. The L2 cache is striped across four banks, each proximate to one near-memory compute engine.  The clock tree diagram for NeCTAr is shown in \autoref{fig:clocking}.  NeCTAr contains two primary clock domains: (1) uncore and system bus and (2) peripheral bus and front bus.  As shown in \autoref{fig:clocking}, the peripherals driving GPIOs, as well as each core, contain separate clock dividers.  The chip implements UART, I2C, PWM, QSPI, and JTAG peripheral interfaces. For access to bulk off-chip memory and debugging, the chip exposes a serialized interface of the TileLink memory protocol, as well as a QSPI port for accessing PSRAM.

\subsection{Near-Memory Compute Engine (NMCE)}

\begin{figure}
    \centering
    \includegraphics[width=0.98\columnwidth]{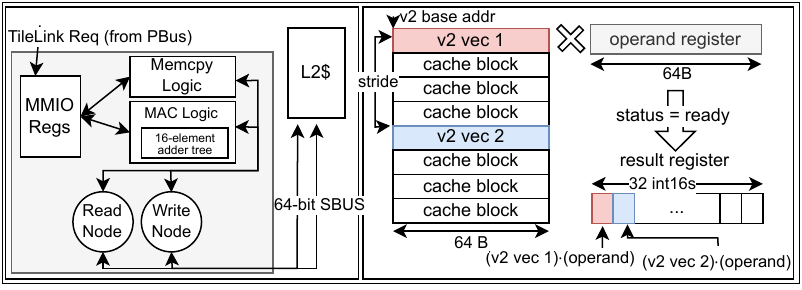}
    \caption{Near-memory compute engine architecture.}
    \label{fig:nmce-arch}
\end{figure}

\begin{figure}
    \centering
    \includegraphics[width=0.9\columnwidth]{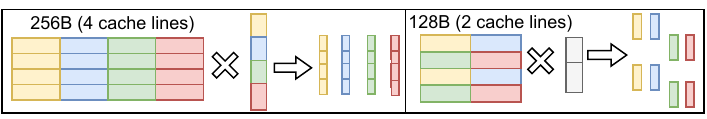}
    \caption{NMCE programming model for 256x4 (left) and 128x4 (right).}
    \label{fig:nmce-pm}
\end{figure}

NeCTAr includes four near-memory compute engines (NMCE), each co-located with one of the four L2 banks. Each module (\autoref{fig:nmce-arch}) supports two operations: multiply-accumulate (MAC) and memory copy (memcpy). The NMCE is programmed through four memory-mapped control registers: (1) \verb|v1Reg|, a 64B register, (2) \verb|v2addr|, a 64b address, (3) a \verb|stride| in bytes for memory accesses and (4) \verb|count|, the number of requested operations (up to 32). The operation computes \verb|count| dot products between int8 vectors and writes each saturated int16 result to a 64B MMIO register. Once the cache line is fetched into the engine, a 64B multiply-accumulate is computed and written in a single cycle. For each dot product, \verb|v1Reg| is used as the first operand, while \verb|v2addr + stride*i| is used as the address of the second operand (\autoref{fig:nmce-arch}). A memory-mapped status register indicates the operation's progress.

With this programming model, the core can parallelize matrix multiplication across the 4 NMCEs for a variety of matrix sizes – examples for a row length of 256B and a row length of 128B are shown in \autoref{fig:nmce-pm}.  The CPU must accumulate across the intermediate results computed by each accelerator.

In \autoref{fig:nmce-workloads}, the NMCE is benchmarked against a single-core software implementation of varying memcpy, MAC, and matmul operations.  Even for a small matrix multiplication of 8 elements by 8 elements, we achieve a 9.66x speedup.  However, as seen in \autoref{fig:eval-results}, with a very large matrix multiplication kernel, using the NMCE leads to a speedup of up to approximately 100x.  Additionally, for a large memcpy of 1MiB, we achieve a speedup of 1.75x. 

\subsection{Sparse Matrix Accelerator}

\begin{figure}
    \centering
    \includegraphics[width=0.8\columnwidth]{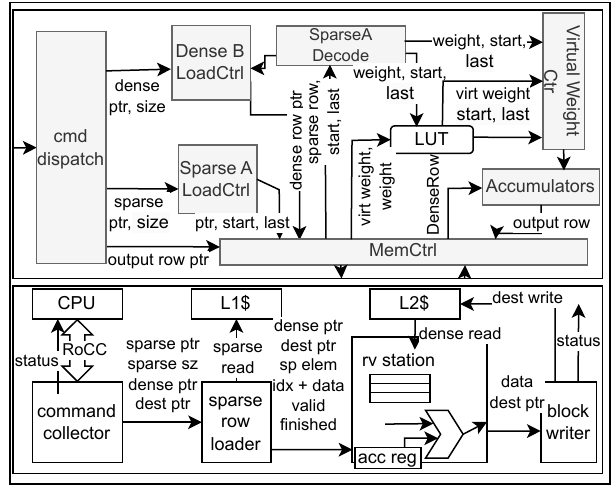}
    \caption{Sparse accelerator architecture.}
    \label{fig:spaccel-arch}
\end{figure}

\begin{figure*}
    \centering
    \includegraphics[width=0.63\textwidth]{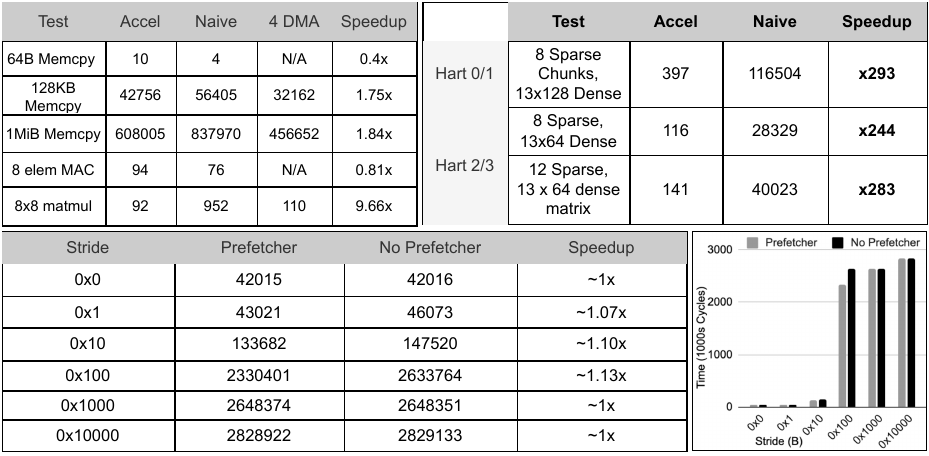}
    \caption{NeCTAr evaluation table.}
    \label{fig:eval-results}
\end{figure*}

The tightly CPU-coupled sparse matrix accelerator (\autoref{fig:spaccel-arch}) advances the architecture in \cite{bearly22-hotchips-poster} by (1) interfacing directly with the L2 cache to improve the memory bandwidth and availability, (2) supporting the translation of virtual addresses and (3) extending support for signed integers. Two of the four accelerators (\autoref{fig:spaccel-arch}, bottom) additionally provide a reservation station to handle L2 requests out-of-order. In \autoref{fig:eval-results}, we benchmark the two variations of the sparse matrix accelerator against a software implementation of sparse-dense matrix multiplication running on a single core. For both variations, we average a 250x speedup on these applications. 

\subsection{Best-Offset Prefetcher}

Our L2 prefetchers are an implementation of Michaud's best-offset prefetcher \cite{best-offset-prefetcher}, a generalization of next-line prefetching that takes into account prefetch timeliness for selecting offset and provides significant speedups over prior sandbox select methods. The algorithm works in sets of learning phases, consisting of several rounds where each candidate offset earns points. At the end of each phase, the offset with the highest score is selected as the best offset, the prefetch offset is updated, and the scores reset to zero. During the chip design phase, the prefetchers were evaluated with the open-source FPGA-accelerated full-system simulation platform FireSim and showed a simulated performance of decreasing Linux boot times by 26\%.  We benchmark the  prefetcher on a simple kernel that does sequential stride accesses, achieving a maximum of 1.13x speedup (\autoref{fig:eval-results}).

\section{Agile Chip Design}

NeCTAr was implemented in Chisel, an open-source hardware description language based on Scala, providing the benefit of object-oriented parameterizable software-defined hardware generation beyond what is possible in Verilog.  Chisel makes use of the FIRRTL and CIRCT intermediate representations and is the basis of the Chipyard framework \cite{chipyard} in which the chip was developed. Chipyard consists of RTL generators such as the Rocket core and Constellation NoC interconnect generator, vastly simplifying the NoC design. Existing Chipyard support for accelerators through the RoCC custom protocol and custom instructions reserved in the RISC-V ISA encoding space allowed teams to quickly integrate accelerator designs with the generated Rocket cores. Chipyard also supports MMIO peripherals through memory-mapped registers, enabling easier integration of our DMA engines. Thanks to Chipyard’s ease of parametrization, the existing open-source modules were extended by the design team with no prior Chipyard, Scala, or VLSI experience to generate test SoCs in less than four weeks. This framework enables for optimizing the complete application performance by both customizing the accelerator architecture and their placement within the multi-core system.  

The Hammer \cite{hammer} physical design tool was used to wrap around commercial EDA tools with a single API and was extended to support the updated PDK for this technology, providing a reusable physical design framework. 

Verification was done through the Chisel unit tests, framework Chisel tests, and integration tests that were all completed by using Chipyard’s Synopsys VCS integration to run C programs. FireSim, the open-source FPGA-accelerated full-system hardware simulation platform, \cite{firesim} was used for realistic performance benchmarking for the prefetcher. The entire design process from conceptualization to verification to tapeout was completed in 15 weeks as a semester-long class.

\section{Evaluation Setup}

\begin{figure}
    \centering
    \includegraphics[width=0.9\columnwidth]{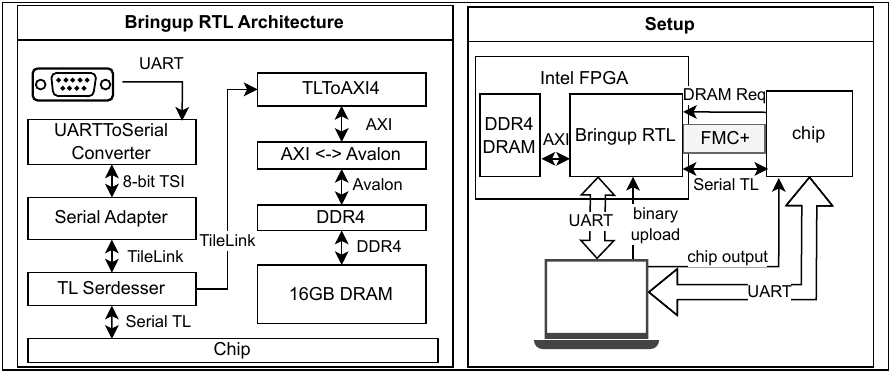}
    \caption{Bringup architecture.}
    \label{fig:bringup-arch}
\end{figure}

The 2mm x 2mm chip in Intel 16 is one of four tiles in a 2x2 array in an interposer package to integrate easily with a custom PCB and FPGA setup for emulation and bring-up. In the main setup, an Intel Agilex FPGA connected to a host through a USB interface sends bring-up commands to the chip across a 560-pin FMC+ connector on the FPGA\@. A custom flow was developed during pre-silicon FPGA emulation by using UART, Tethered Serial Interface (TSI), and the serialized TileLink interface on the chip to enable communication. Because serialized TileLink allows read and write commands to any memory address from off-chip without involving the cores, this protocol is a convenient bring-up method. The chip accesses the DDR4 memory of the FPGA through the same FMC and serialized TileLink interface. Before the chip tapeout, the design was prototyped on an FPGA by using the setup shown in (\autoref{fig:bringup-arch}, right) by replacing the actual chip with another FPGA running the SoC’s RTL\@. 

\section{Measurement and Results}

For evaluation, the chip is mounted on the PCB provided by Intel and connected to the FPGA through FMC+.  Upon completion of the workload, the chip signals through GPIO to an ESP32 microcontroller, which is used to measure the timing of each benchmark and reset the chip. The chip operates from the nominal 850mV supply down to 550mV\@.  \autoref{fig:nmce-workloads} shows the evaluation of NeCTAr using NMCE vs. multi-core software for a very large matrix multiply kernel.  

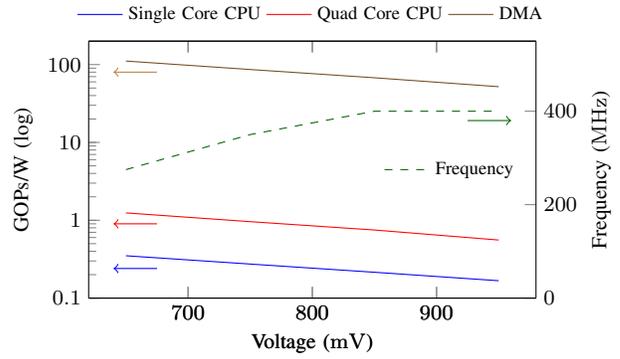
\begin{figure}
    \centering
    \begin{tikzpicture}
    \footnotesize
    \pgfplotsset{set layers,width=0.85\columnwidth,height=5cm}
    \begin{axis}[
        axis y line*=right,
        ymin=0,
        ymax=550,
        xlabel={Voltage (\si{\milli\volt})},
        ylabel={Frequency (\si{\mega\hertz})},
        legend style={font={\scriptsize},at={(0.98,0.5)},anchor=east,draw=none},
    ]
        \addplot+[mark=none,dashed,color=green!40!black] table[col sep=comma,x=V,y=freq] {shmoo.csv};
        \addlegendentry{Frequency}
        \draw[->,green!40!black] (axis cs:925,380) -- (axis cs:960,380);
    \end{axis}
    \begin{semilogyaxis}[
        axis y line*=left,
        ymin=0.1,
        ymax=200,
        log ticks with fixed point,
        xlabel={Voltage (\si{\milli\volt})},
        ylabel={GOPs/W (log)},
        legend columns=3,
        legend style={font={\scriptsize},at={(0.5,1.03)},anchor=south,draw=none},
    ]
        \addplot+[mark=none] table[col sep=comma,x=V,y=singlecore] {shmoo.csv};
        \addlegendentry{Single Core CPU}
        \addplot+[mark=none] table[col sep=comma,x=V,y=quadcore] {shmoo.csv};
        \addlegendentry{Quad Core CPU}
        \addplot+[mark=none] table[col sep=comma,x=V,y=dma] {shmoo.csv};
        \addlegendentry{DMA}
        \draw[->,red] (axis cs:675,0.9) -- (axis cs:640,0.9);
        \draw[->,blue] (axis cs:675,0.24) -- (axis cs:640,0.24);
        \draw[->,brown] (axis cs:675,80) -- (axis cs:640,80);
    \end{semilogyaxis}
    \end{tikzpicture}
    \vspace{-2ex}
    \caption{Efficiency and frequency curves.}
    \label{fig:shmoo}
\end{figure}

At the nominal voltage, the maximum clock frequency is 400MHz, and the highest efficiency is 132 GOPs/W\@. The highest performance obtained is 6.02 GOPs (where INT8 is a unit operation), an approximately 100x speedup compared to the multi-core 56.6 MOPs. A  Shmoo plot for the chip, as well as performance curves comparing single-core, multi-core, and NMCE performance, are shown in \autoref{fig:shmoo}.  For this multiplication kernel, using the  NMCE shows a 100x improvement over distributing the matrix multiplication among four cores in software.

\subsection{End-to-End Application}

\begin{table}
    \caption{Comparison chart.}
    \label{fig:comparison}
    \vspace{-6pt}
    \centering
    \begingroup
    \scriptsize
    \setlength{\tabcolsep}{1pt}
    \begin{tabular}{>{\itshape}c *{4}{c}}
        Name & NeCTAr & Chen CNC \cite{chen-cnc} & Rovinski \cite{rovinski} & Thestral \cite{thestral} \\ \midrule
        Technology & Intel 16nm & Intel 4 & 16nm FinFet & GF 22FDX \\
        Die Area & \SI{4}{\milli\meter\squared} & \SI{1.92}{\milli\meter\squared} & \SI{15.25}{\milli\meter\squared} & \SI{1}{\milli\meter\squared} \\
        Int Precision & INT8 & INT8 & INT32 & INT32 \\
        Voltage & 0.55--0.85V & 0.6--0.82V & 0.6--0.98V & 0.6--0.9V \\
        Power & 7.3--171 \si{\milli\watt} & N/A & 7.47 \si{\watt} & N/A \\
        Max Freq & 400 MHz & 1.15 GHz & 1.4 GHz & 910 MHz \\
        Peak Eff. & 132 GOPs/W & 285 GOPs/W & 93.04 GOPs/W & N/A \\
    \end{tabular}
    \endgroup
\end{table}

To demonstrate the effectiveness of NeCTAr on end-to-end applications, we run a 1.7M ReLU-Llama trained on the TinyStories dataset \cite{eldan2023tinystories} and on an on-chip synthetic dense matrix workload.  The ReLU-Llama model is modified to induce activation sparsity to halve weight reads from off-chip \cite{mirzadeh2024relu}. By taking advantage of activation sparsity, the model can reduce the number of off-chip memory accesses. A comparison with recent research inference SoCs is in \autoref{fig:comparison}.  As seen from \autoref{fig:nmce-workloads}, NeCTAr can achieve up to 45.4 infs/s/W and 1.28 infs/s.  The 1.7M model doesn't fit into the L2 cache, so NeCTAr communicates with external DRAM through a narrow, single-wire  serialized TileLink interface.  

\begin{table}
    \caption{End-to-end application performance.}
    \label{fig:nmce-workloads}
    \vspace{-6pt}
    \centering
    \begingroup
    \footnotesize
    \setlength{\tabcolsep}{0.5em}
        \begin{tabular}{*{3}{c}}
            \multicolumn{3}{c}{\textit{Matrix Multiplication}} \\
            Single-core & Multi-core & NMCE \\ \midrule
            0.350 GOPs/W & 1.24 GOPs/W & 132 GOPs/W \\
            18.7 MOPs & 56.6 MOPs & 6.02 GOPs \\
        \end{tabular}
    \\[2ex]
        \begin{tabular}{*{3}{c}}
            \multicolumn{3}{c}{\textit{1.7B LLAMA}} \\
            Single-core & Multi-core & NMCE \\ \midrule
            39.0 infs/s/W & 40.0 infs/s/W & 45.4 infs/s/W \\
            1.19 infs/s & 1.25 ints/s & 1.28 infs/s \\
        \end{tabular}
    \endgroup
\end{table}

\section{Conclusion}

This paper demonstrates use of a framework that enables agile chip design during the course of 15 weeks.  NeCTAr is a multi-core heterogenous SoC that uses custom near-memory dense matrix and CPU-coupled sparse matrix accelerators to efficiently run complex applications, including a 1.7M LLM inference model. By reusing the framework, an SoC can be generated to target evolving DNN targets.

\section*{Acknowledgment}

We thank the Intel University Shuttle Program for donating the chip fabrication and packaging; NSF CCRI ENS Chipyard Award \#2016662; Scalable Asymmetric Lifecycle Engagement (SCALE); Apple’s New Silicon Initiative support. We acknowledge Bryan Casper, Matt Rebsom, and Nancy Robinson from Intel for support, SLICE and BWRC students, staff, and member companies. This work is a collaboration of dozens of UC Berkeley students from the Spring Tapeout class and bring-up students from the Fall Bring-Up class. We thank the many mentors that made the efforts possible.

\bibliographystyle{IEEEtran}
\bibliography{IEEEabrv,./bibdb}

\end{document}